\documentclass[10pt,conference]{IEEEtran}
\IEEEoverridecommandlockouts
\usepackage{cite}
\usepackage{amsmath,amssymb,amsfonts}
\usepackage[acronyms]{glossaries}
\usepackage{algorithmic}
\usepackage{graphicx}
\usepackage{subcaption}
\usepackage{textcomp}
\usepackage{xcolor}
\usepackage{footnote}
\usepackage{soul}

\newcommand {\Define} {\stackrel {\Delta} {=}  }
\makeatletter
\newcommand{\ostar}{\mathbin{\mathpalette\make@circled *}}
\newcommand{\make@circled}[2]{%
  \ooalign{$\m@th#1\smallbigcirc{#1}$\cr\hidewidth$\m@th#1#2$\hidewidth\cr}%
}
\newcommand{\smallbigcirc}[1]{%
  \vcenter{\hbox{\scalebox{0.77778}{$\m@th#1\bigcirc$}}}%
}
\makeatother

\def\BibTeX{{\rm B\kern-.05em{\sc i\kern-.025em b}\kern-.08em
    T\kern-.1667em\lower.7ex\hbox{E}\kern-.125emX}}

\newacronym{sdr}{SDR}{software-defined radio}
\newacronym{iq}{IQ}{in-phase and quadrature}
\newacronym{if}{IF}{intermediate frequency}
\newacronym{lna}{LNA}{low-noise amplifier}
\newacronym{rf}{RF}{radio frequency}
\newacronym{6g}{6G}{sixth generation}
\newacronym{vr}{VR}{virtual reality}
\newacronym{adc}{ADC}{analog to digital converter}
\newacronym{dac}{DAC}{digital to analog converter}
\newacronym{awg}{AWG}{arbitrary waveform generator}
\newacronym{dso}{DSO}{digital storage oscilloscope}
\newacronym{psg}{PSG}{performance signal generator}
\newacronym{thz}{THz}{terahertz}
\newacronym{sub-thz}{sub-THz}{sub-terahertz}
\newacronym{fft}{FFT}{fast Fourier transform}
\newacronym{css}{CSS}{chirp spread spectrum}
\newacronym{ofdm}{OFDM}{Orthogonal Frequency Division Multiplexing}
\newacronym{lo}{LO}{local oscillator}
\newacronym{otfs}{OTFS}{Orthogonal Time Frequency Space}
\newacronym{papr}{PAPR}{peak-to-average power ratio}
\newacronym{mmse}{MMSE}{minimum mean squared error}
\newacronym{ber}{BER}{bit error rate}
\newacronym{vdi}{VDI}{Virginia Diodes Inc.}
\newacronym{awgn}{AWGN}{additive white Gaussian noise}
\newacronym{dft}{DFT}{discrete Fourier transform}
\newacronym{psk}{PSK}{phase-shift keying}
\newacronym{qam}{QAM}{quadrature amplitude modulation}
\newacronym{evm}{EVM}{error vector magnitude}
\newacronym{bpsk}{BPSK}{binary phase-shift keying}
\newacronym{qpsk}{QPSK}{quadrature phase-shift keying}
\newacronym{ifft}{IFFT}{inverse fast Fourier transform}
\newacronym{sfft}{SFFT}{symplectic finite Fourier transform}
\newacronym{isfft}{ISFFT}{inverse symplectic finite Fourier transform}
\newacronym{idft}{IDFT}{inverse discrete Fourier transform}
\newacronym{cfo}{CFO}{carrier frequency offset}
\newacronym{cpo}{CPO}{carrier phase offset}
\newacronym{mixamc}{MixAMC}{mixer/amplifier/multiplier-chain}
\newacronym{usrp}{USRP}{universal software radio peripheral}
\newacronym{dd}{DD}{delay-Doppler}
\newacronym{tx}{Tx}{transmitter}
\newacronym{rx}{Rx}{receiver}
\newacronym{los}{LOS}{line-of-sight}
\newacronym{snr}{SNR}{signal-to-noise ratio}
\newacronym{td}{TD}{time domain}
\newacronym{2d}{2D}{two dimensional}
\newacronym{lpf}{LPF}{low-pass filter}
\newacronym{dzt}{DZT}{discrete time Zak transform}
\newacronym{ml}{ML}{maximum likelihood}
\newacronym{isac}{ISAC}{integrated sensing and communications}
\newacronym{aiml}{AI/ML}{artificial intelligence/machine learning}
\newacronym{io}{I/O}{input-output}

\begin{document}


\title{Over-the-Air Transmission of Zak-\gls{otfs} with Spread Pilots on Sub-THz Communications Testbed \vspace{-5mm}
\thanks{This work was supported by the NSF (2342690, 2342690, and 214821), the AFOSR (FA8750-20-2-0504 and FA9550-23-1-0249), and by funds from federal agency and industry partners as specified in the Resilient \& Intelligent NextG Systems (RINGS) program. \\
\par \textbf{This work may be submitted to the IEEE for possible publication. Copyright
may be transferred without notice, after which this version may no longer be
accessible.}}
}



 \author{
   \IEEEauthorblockN{Claire Parisi\IEEEauthorrefmark{2}, Venkatesh Khammammetti\IEEEauthorrefmark{1}, Robert Calderbank\IEEEauthorrefmark{1}, and Lauren Huie\IEEEauthorrefmark{2}}
   \IEEEauthorblockA{\IEEEauthorrefmark{1}\textit{Electrical and Computer Engineering Department, Duke University}, Durham, NC, USA \\ \IEEEauthorrefmark{2}\textit{Information Directorate, Air Force Research Laboratory}, Rome, NY, USA  \\ claire.parisi@us.af.mil, venkatesh.khammammetti@duke.edu, robert.calderbank@duke.edu, and lauren.huie-seversky@us.af.mil\vspace{-1.2em}
   }
}

\maketitle

\begin{abstract}
Looking towards 6G wireless systems, frequency bands like the \gls{sub-thz} band (100~GHz - 300~GHz) are gaining traction for their promises of large available swaths of bandwidth to support the ever-growing data demands. However, challenges with harsh channel conditions and hardware nonlinearities in the \gls{sub-thz} band require robust communication techniques with favorable properties, such as good spectral efficiency and low \gls{papr}.  Recently, \gls{otfs} and its variants have garnered significant attention for their performance in severe conditions (like high delay and Doppler), making it a promising candidate for future communications. In this work, we implement Zak-\gls{otfs} for the over-the-air experiments with traditional point pilots and the new spread pilots. Notably, we design our spread-pilot waveforms with communications and sensing coexisting in the same radio resources. We define the system model and the signal design for integration onto our state-of-the-art \gls{sub-thz} wireless testbed. We show successful data transmission over-the-air at 140~GHz and 240~GHz in a variety of \gls{snr} conditions. In addition, we demonstrate \gls{isac} capabilities and show \gls{papr} improvement of over 5~dB with spread pilots compared to point pilots. \end{abstract}
\glsresetall


\section{Introduction}
To meet the demand for ultra-high data rates in next-generation wireless systems, attention has turned towards emerging spectral, such as the \gls{sub-thz} band (100~GHz - 300~GHz), with large contiguous swaths of bandwidth capable of supporting high-data, wideband applications. While systems in this band have already shown great promise~\cite{Elayan2020,Jiang2024}, these frequencies come with a unique set of challenges, including higher path losses, Doppler spread, and suceptibilty to phase noise and \gls{papr}~\cite{Akyildiz2022,Lee2020}. \gls{otfs}, as presented in~\cite{Hadani2017}, is an emerging waveform candidate designed specifically to overcome harsh channel conditions through strategic design and resource allocation in the \gls{dd} domain. Inherent properties grant \gls{otfs} significant advantages in high-mobility scenarios and challenging propagation environments, such as the \gls{sub-thz} regime, by making the \gls{io} relationship predictable. Previous studies in \cite{Taraboush2022,Parisi2024} show that \gls{otfs} and spread-variations on \gls{otfs} show potential in the context of \gls{sub-thz} band hardware and channel challenges. Specifically, \gls{otfs} often shows improved performance over \gls{ofdm}, a technique that is widely used in current 5G and LTE systems, in a variety of studies and scenarios~\cite{Gaudio2022,Liu2025,Wiffen2018}. However, \gls{otfs} suffers from high \gls{papr}, making spread variations attractive candidates (like spread pilots, which reduce \gls{papr} through filtering)~\cite{ubadah2024}. Zak-\gls{otfs} is a new variant of \gls{otfs} \cite{Saif2_base} that is designed to make the I/O relation predictable even when the propagation environment is harsh, and it can also reduce the \gls{papr} \cite{ubadah2024}. In addition, future wireless systems in 6G and beyond are projected to support \gls{aiml} and \gls{isac} capabilities~\cite{Jiang2024}, making \gls{otfs} an even more advantageous technique since it generates a time-invariant channel representation (better for training \gls{aiml} models) and supports the overlay of sensing and communications resources in the same frame. 

As such, in this paper, we focus our attention on the implementation of Zak-\gls{otfs} system with both point and spread pilots on practical \gls{sub-thz} hardware to showcase feasibility for next-generation communications systems. We design \gls{td} \gls{otfs} packets in software using our Zak-\gls{otfs} transmitter and receiver models. We model two different packets, one using point pilots (where the data frame is proceeded by the pilot frame) and a second for \gls{isac} (where the spread pilots and data are overlayed within the same frame). From there, we integrate our waveforms onto \gls{usrp} B210 radios connected to \gls{sub-thz} frontends to transmit both waveforms over-the-air at 140~GHz and 240~GHz. We successfully demonstrate data transmission and sensing capabilities on the \gls{sub-thz} testing setup. We see promising performance of Zak-\gls{otfs} in a variety of \gls{snr} scenarios and demonstrate preliminary \gls{isac} capabilities in addition to reduced-\gls{papr} with spread pilots.

The remainder of this paper is organized as follows: in Section~\ref{section:Transceiver Design}, we describe the system design details of Zak-\gls{otfs}; in
Section~\ref{section: testbed overview} we overview our \gls{sub-thz} testing hardware setup; in Section~\ref{section:Otfs implementation}, we discuss how we integrate our \gls{otfs} design onto the testbed; in
Section~\ref{section:results}, we showcase the results of wireless transmission of Zak-\gls{otfs} with point pilots and spread pilots; lastly, we conclude in Section~\ref{section:conclusion}

\section{\gls{otfs} System Design}
\label{section:Transceiver Design}
For our \gls{otfs} transceiver design, we choose Zak-\gls{otfs} (also referred to as \gls{otfs} 2.0) \cite{Saif2_base} as our model. Zak-\gls{otfs} transforms the \gls{2d} \gls{dd} signal directly into the \gls{td}  using the Zak transform, which is proven to be more robust to channel delay and Doppler spreads, more effective in predicting the \gls{io} relation and has good spectral efficiency performance compared to other versions of \gls{otfs} (like the most studied MC-\gls{otfs}) \cite{Saif2_chan}. When implementing Zak-\gls{otfs} systems with a point pilot, we introduce a guard band around the pilot location, within which no data is transmitted to get an efficient channel prediction to equalize on the data. However, this decreases the overall spectral efficiency and also has has high \gls{papr}, because of the point pilots.  To overcome these issues, the spread-pilot-based channel prediction technique is introduced in \cite{ubadah2024}. The remainder of this section describes the Zak-OTFS system (transmitter and receiver) used for the over-the-air experimentation on the \gls{sub-thz} bands and also explains the point pilot and spread pilot. 
\subsection{Zak-\gls{otfs} Transmitter}
\label{subsection:zak_tx}
\begin{figure}[htb]
    \centering
    \includegraphics[width=\linewidth]{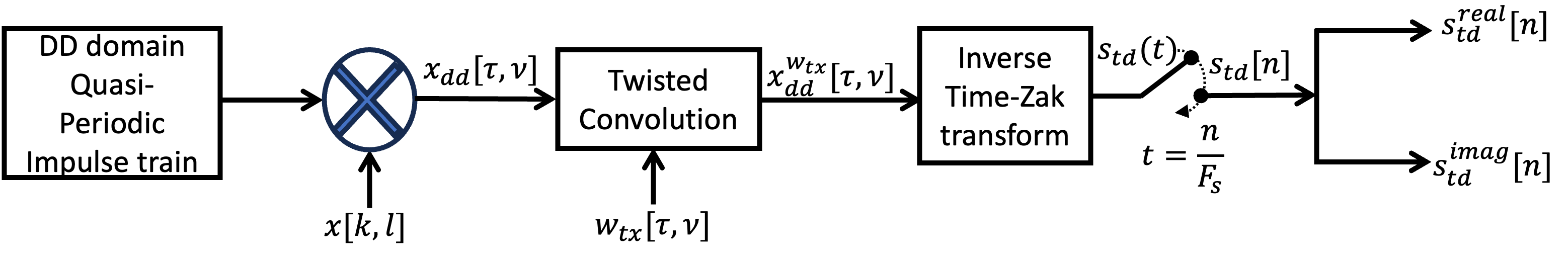}
    \caption{Zak-OTFS baseband Modulation}
    \label{fig:bbmod}
\end{figure}\vspace{-1mm}
Consider the fundamental periods of the \gls{dd} domain to be $\tau_p$ along delay and $\nu_p$ along the Doppler domains, ($\tau_p \cdot \nu_p = 1$). The delay and Doppler periods are sub-divided into $M$ and $N$ equal parts, respectively, such that each \gls{dd} resolution is $\left(\frac{\tau_p}{M}=\frac{1}{B},\frac{\nu_p}{N}=\frac{1}{T}\right)$, where $B$ and $T$ are the bandwidth and time duration of the Zak-OTFS signal. The $MN$ information symbols $x[k,l]$, $k=0,1,...,M-1$, $l=0,1,...,N-1$ are mapped onto the $MN$ \gls{dd} resolutions in the fundamental domain such that $x[k,l] = x[k\tau_p/M,l\nu_p/N]$. 
From \cite{Saif2_base} and the properties of the Zak transform, the \gls{dd} input signal to the inverse Zak transform should be quasi-periodic\footnote{quasi-periodic means periodic with a phase shift} in nature. From \cite{Saif2_base}, the quasi-periodic continuous \gls{dd} domain signal is defined as \vspace{-3mm}
        \begin{gather}
          x_{_{\mbox{\scriptsize{dd}}}}(\tau,\nu)  \Define  \sum\limits_{k,l \in {\mathbb Z}} x_{_{\mbox{\scriptsize{dd}}}}[k,l] \, \delta(\tau - k/B) \, \delta(\nu - l/T), \nonumber \\ 
x_{_{\mbox{\scriptsize{dd}}}}[k+n M, l+m N]   \Define  x[k, l] e^{j 2 \pi n \frac{l}{N}}, \textup{ } m, n \in \mathbb{Z}.     \label{eqn1}
    \end{gather}
From Fig.\ref{fig:bbmod}, the output of the twisted convolution is given by \vspace{-2mm}
\begin{eqnarray}
\label{twistconv}
x_{_{\mbox{\scriptsize{dd}}}}^{w_{tx}}(\tau,\nu) \hspace{-3mm} & = & \hspace{-3mm}w_{tx}(\tau, \nu) \, *_{\sigma} \, x_{_{\mbox{\scriptsize{dd}}}}(\tau,\nu) \nonumber \\
\hspace{-8mm}&=&\hspace{-3mm} 
 \iint w_{tx}\left(\tau^{\prime}, \nu^{\prime}\right) x_{dd}\left(\tau-\tau^{\prime}, \nu-\nu^{\prime}\right) \nonumber \\
 & & \hspace{25mm} e^{j 2 \pi \nu^{\prime}\left(\tau-\tau^{\prime}\right)} d \tau^{\prime} d \nu^{\prime}
\end{eqnarray}
In this paper, we consider 2D \gls{dd} `sinc' filter, $w_{tx}(\tau,\nu) \Define \sqrt{B T} \operatorname{sinc}(B \tau) \operatorname{sinc}(T \nu)$ (refer \cite{Saif2_chan, Jinu_Gauss, Das_Gauss} for other \gls{dd} filters). Substituting (\ref{eqn1}) and $w_{tx}(\tau,\nu)$ in (\ref{twistconv}) gives \vspace{-2mm} 
\begin{eqnarray}
\label{x_wtx_dd}
x_{_{\mbox{\scriptsize{dd}}}}^{w_{tx}}(\tau,\nu) &=& \sqrt{B T} \sum\limits_{k,l \in {\mathbb Z}} x_{_{\mbox{\scriptsize{dd}}}}[k,l] \,\ \operatorname{sinc}\big(T (\nu-\frac{l}{T})\big) \nonumber \\
& & \hspace{9mm} \operatorname{sinc}\big(B (\tau-\frac{k}{B})\big) e^{j2\pi\left(\nu-\frac{l}{T}\right)\frac{k}{B}}  \, 
\end{eqnarray}
Applying the inverse time Zak transform  on $x_{_{\mbox{\scriptsize{dd}}}}^{w_{tx}}(\tau,\nu)$ (see Fig. \ref{fig:bbmod}) results in a physical \gls{td} signal $s_{_{\mbox{\scriptsize{td}}}}(t)$ \cite{Saif2_base}, i.e.,\vspace{-2mm}
\begin{eqnarray}
\label{s_td}
    s_{_{\mbox{\scriptsize{td}}}}(t) &=& \sqrt{\tau_{p}}\int_{0}^{\nu_{p}} x_{_{\mbox{\scriptsize{dd}}}}^{w_{tx}}(t,\nu) d\nu
\end{eqnarray}
Substituting (\ref{x_wtx_dd}) in (\ref{s_td}) (see \cite{saif_book} for the detailed derivation) gives, \vspace{-6mm}
\begin{eqnarray}
\label{s_td_final}
s_{_{\mbox{\scriptsize{td}}}}(t) &=& \sqrt{\frac{B \tau_{p}}{T}} \sum_{k=0}^{M-1} \sum_{l=0}^{N-1} x[k, l]\left[\sum_{n=-\frac{N}{2}}^{\frac{N}{2}-1} e^{j 2 \pi \frac{n l}{N}} \right. \nonumber \\
& & \hspace{12mm} \left. \operatorname{sinc}\left(B\left(t-n \tau_{p}-\frac{k \tau_{p}}{M}\right)\right)\right]
\end{eqnarray}








Now the complex \gls{td} signal $s_{_{\mbox{\scriptsize{td}}}}(t)$ is  sampled at a sampling rate $F_s \geq B$, where $B$ is the Nyquist rate. Therefore, the $q$-th sample obtained by sampling $s_{_{\mbox{\scriptsize{td}}}}(t)$ at $t=q/F_s$ is given by \vspace{-3mm}
\begin{eqnarray}
    \label{s_td_samples}
    s_{_{\mbox{\scriptsize{td}}}}[q] &\triangleq& s_{_{\mbox{\scriptsize{td}}}}\left(t=q / F_{s}\right)=\sqrt{\frac{B \tau_{p}}{T}} \sum_{k=0}^{M-1} \sum_{l=0}^{N-1} x[k, l] \nonumber \\
    & & \hspace{-3mm}\left[\sum_{n=-\frac{N}{2}}^{\frac{N}{2}-1} e^{j 2 \pi \frac{n l}{N}} \operatorname{sinc}\left(B\left(\frac{q}{F_{s}}-n \tau_{p}-\frac{k \tau_{p}}{M}\right)\right)\right]\ \   
\end{eqnarray}
Since $s_{_{\mbox{\scriptsize{td}}}}(t)$ has time duration $T$ and lies in the interval $[-T / 2, T / 2), \,\ s_{_{\mbox{\scriptsize{td}}}}[q]$ takes non-zero values only for $q=-\frac{F_{s} T}{2}, \cdots, \frac{F_{s} T}{2}$, i.e., roughly $F_{s} T$ samples. 
\subsection{Zak-\gls{otfs} Receiver} \vspace{-2mm}
\label{subsection:receiver}
\begin{figure}[htb]
    \centering
    \includegraphics[width=0.8\linewidth]{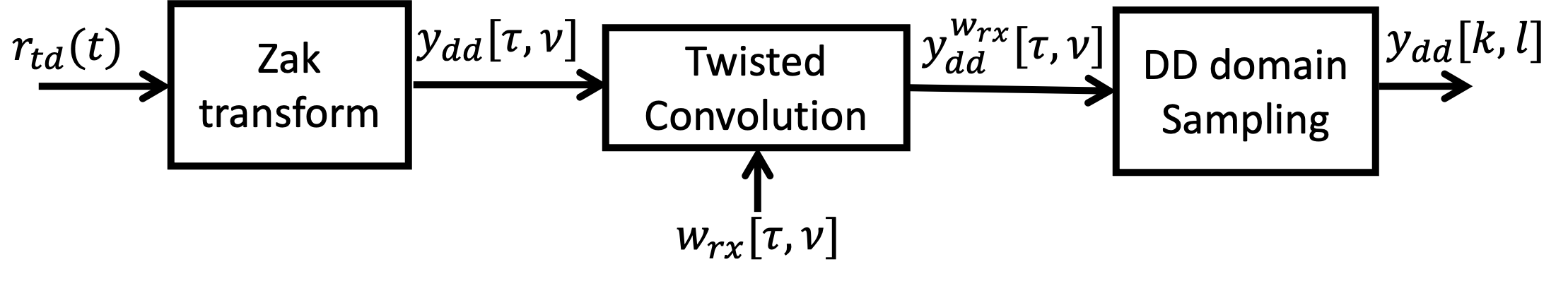}
    \caption{Zak-OTFS baseband Demodulation}
    \label{fig:bb_demod}
\end{figure} \vspace{-2mm}
At the receiver, a time Zak transform is applied to the received baseband \gls{td} signal $r_{_{\mbox{\scriptsize{td}}}}(t)$ to obtain a \gls{2d} \gls{dd} signal $y_{_{\mbox{\scriptsize{dd}}}}(\tau,\nu)$ (See Fig. \ref{fig:bb_demod}).  This signal ($y_{_{\mbox{\scriptsize{dd}}}}(\tau,\nu)$) is twisted convolved with the \gls{dd} receiver pulse shaping filter $w_{rx}(\tau,\nu)$ to get $y_{_{\mbox{\scriptsize{dd}}}}^{w_{rx}}$. The \gls{dd} sampled signal $y_{_{\mbox{\scriptsize{dd}}}}[k,l]$ is obtained by sampling $y_{_{\mbox{\scriptsize{dd}}}}^{w_{rx}}(\tau,\nu)$ at $\tau=k\frac{\tau_p}{M}, \nu = l\frac{\nu_p}{N}$. The entire process of converting the received baseband signal $r_{_{\mbox{\scriptsize{td}}}}(t)$ to \gls{dd} a sampled signal $y_{_{\mbox{\scriptsize{dd}}}}[k,l]$ (see Fig. \ref{fig:bb_demod}) can be done in two steps.
\begin{enumerate}
  \item In the first step,the received baseband \gls{td} signal $r_{_{\mbox{\scriptsize{td}}}}(t)$ is passed through an ideal \gls{lpf} and the filtered output is given by \vspace{-4mm}
\end{enumerate}
\begin{eqnarray}
    \label{y_lpf}
    y(t)=r_{_{\mbox{\scriptsize{td}}}}(t) *(B \operatorname{sinc}(B t))\hspace{-1mm}=\hspace{-1mm}B \hspace{-1mm}\int \hspace{-1mm} \operatorname{sinc}(B \tau) r_{_{\mbox{\scriptsize{td}}}}(t-\tau) d \tau
    \vspace{-2mm}
\end{eqnarray} 
Now limiting the \gls{td} signal $y(t)$ to the TD interval $[-T / 2, T / 2]$ collect the samples at sampling rate $B$ such that
$y[n]=y\left(t=\frac{n}{B}\right)$.
\begin{enumerate}
  \setcounter{enumi}{1}
  \item In the second step, apply \gls{dzt} to convert the discrete-time signal $y[n]$ to the discrete-\gls{dd} domain signal $y_{_{\mbox{\scriptsize{dd}}}}[k, l]$, i.e., \vspace{3mm}\newline
\end{enumerate}
\begin{eqnarray}
\label{y_dd_rx}
\begin{aligned}
y_{_{\mbox{\scriptsize{dd}}}}[k, l]= & \sum_{n=-\frac{N-1}{2}}^{\frac{N-1}{2}} y[k+n M] e^{-j 2 \pi \frac{n l}{N}}, \\
& k=0,1, \cdots, M-1, l=0,1, \cdots, N-1 .
\end{aligned}
\end{eqnarray}\vspace{-1mm}
Note that step one of this two-step process is only valid if the  receiver pulse shaping filter is sinc i.e., $w_{rx}(\tau,\nu) = \sqrt{B T} \operatorname{sinc}(B \tau) \operatorname{sinc}(T \nu)$.
\par The \gls{dd} received signal $y_{_{\mbox{\scriptsize{dd}}}}[k, l]$ is the twisted convolution of the effective DD domain channel filter $h_{\mathrm{eff}}[k, l]$ with the input, and  the channel filter $h_{\mathrm{eff}}[k, l]$ is formally defined in [13]. Using $y_{_{\mbox{\scriptsize{dd}}}}[k, l]$, we predict the effective discrete \gls{dd} channel filter $h_{\mathrm{eff}}[k, l]$. The \gls{ml} estimate of $h_{\mathrm{eff}}[k, l]$ is then given by the cross-ambiguity between the received pilot signal $y_{_{\mbox{\scriptsize{dd}}}}[k, l]$ and the transmitted pilot signal $x_{\mathrm{i}, \mathrm{dd}}[k, l]$ (refer \cite{ubadah2024} for more details), i.e., \vspace{-2mm}
\begin{eqnarray}
    \label{cross_am}
    \begin{aligned}
& \widehat{h}_{\mathrm{eff}}[k, l]=A_{y_s, x_s}[k, l] \\
& =\sum_{k^{\prime}=0}^{M-1} \sum_{l^{\prime}=0}^{N-1} y_{_{\mbox{\scriptsize{dd}}}}\left[k^{\prime}, l^{\prime}\right] x_{\mathrm{i}, \mathrm{dd}}^*\left[k^{\prime}-k, l^{\prime}-l\right] e^{-j 2 \pi \frac{l\left(k^{\prime}-k\right)}{M N}},
\end{aligned} 
\end{eqnarray}
for $(k, l) \in \mathcal{S}$, where $\mathcal{S}$ is the support set over which the \gls{dd} channel spreads and $x_{\mathrm{i}, \mathrm{dd}}[k,l]$ can be point or spread pilot signal used during transmission. After estimating, $\widehat{h}_{\mathrm{eff}}[k, l] \,\ \forall k,l$ we generate the predicted channel matrix, (as described in \cite{Saif2_chan}), and perform \gls{mmse} equalization to estimate the transmitted information symbols. \vspace{-2mm}
 \subsection{Pilot signal} \vspace{-1mm}
 We predict the I/O relation from the pilot response (the data complicates prediction). For the experiments performed in this paper, we transmitted two types of pilots in the \gls{dd} domain: $\left.1\right)$ point pilot and $\left.2\right)$ spread pilot.  
\subsubsection{Point Pilot}
A symbol located at $(k_p,l_p)$ in the \gls{dd} domain. i.e., $x_{_{\mbox{\scriptsize{p}}}}[k, l]= \delta[k-k_p]\delta[l-l_p]$.
From the definition of quasi-peridocity (see (\ref{eqn1})), the discrete quasi-periodic \gls{dd} domain point pilot is given by \vspace{-2mm}
\begin{eqnarray}
    \label{Quasi_ppilot}
    x_{_{\mbox{\scriptsize{p,dd}}}}[k,l] = \hspace{-2mm}\sum_{n,m \in \mathbb{Z}} e^{j2\pi\frac{nl}{N}}\delta[k-k_p-nM]\delta[l-l_p-mN]
\end{eqnarray}
\subsubsection{Spread Pilot}
The \gls{dd} domain signal obtained by $MN$-periodic twisted convolution of a $MN$-periodic discrete \gls{dd} filter $w[k,l]$ with the discrete quasi-periodic signal $x_{_{\mbox{\scriptsize{p,dd}}}}[k,l]$, i.e., \vspace{-4mm}
\begin{eqnarray}
    \label{spread_pilot}
    x_{_{\mbox{\scriptsize{s,dd}}}}[k,l] = w[k,l] \ostar_{\sigma} x_{_{\mbox{\scriptsize{p,dd}}}}[k,l]
\end{eqnarray} \vspace{-5mm}
\begin{figure}[htb]
    \centering
    \includegraphics[width=\linewidth]{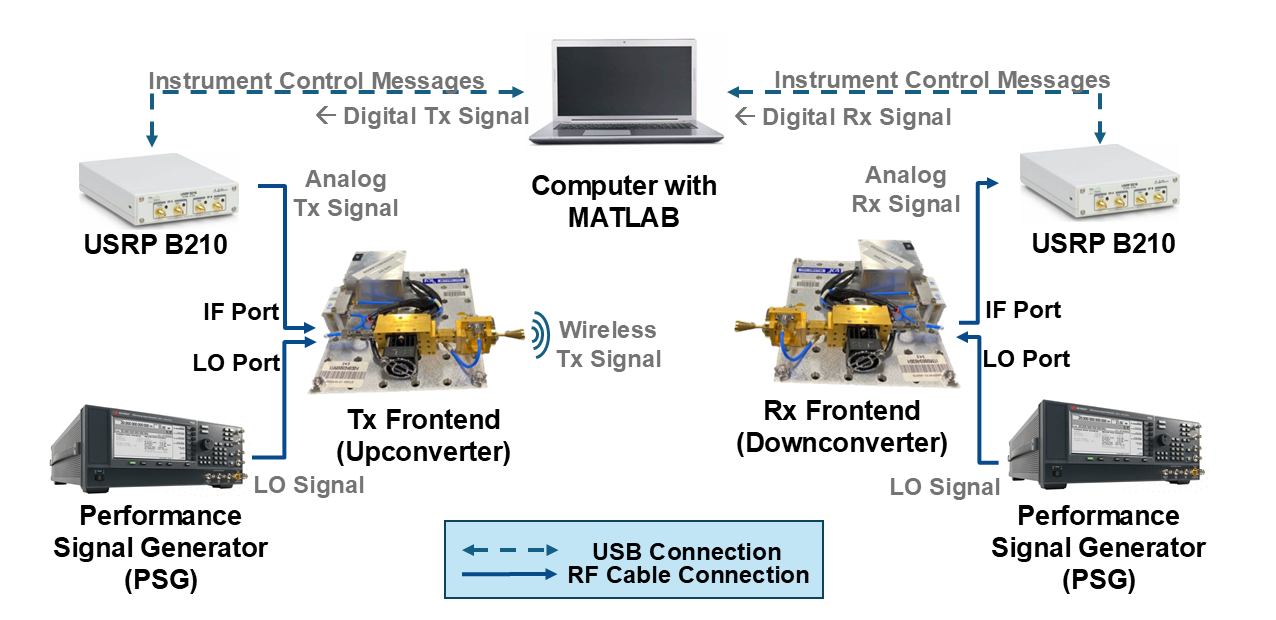}
    \caption{Sub-THz testbed system diagram}
    \label{fig:connection diagram}
\end{figure}
\vspace{4mm}where $\ostar_{\sigma}$ denotes the $MN$-periodic twisted convolution defined in \cite{ubadah2024}. 
The $MN$-periodic discrete \gls{dd} filter $w[k,l]$ is given by \vspace{-3mm}
\begin{eqnarray}
    \label{spread_filter}
    w[k,l] = \frac{1}{MN}e^{j2\pi\frac{u\left(k^2+l^2\right)}{MN}}, \,\ \forall \,\  k,l \in \mathbb{Z},
\end{eqnarray}
where $u \in \mathbb{Z}$ is the slope parameter. Substituting (\ref{spread_filter}) in (\ref{spread_pilot}) gives \vspace{-4mm}
\begin{eqnarray}
\label{spreadpilot_eqn}
x_{_{\mbox{\scriptsize{s,dd}}}}[k,l]
 &=&  \sum_{n=0}^{N-1} \sum_{m=0}^{M-1} e^{j 2 \pi n \frac{\left(l_p+m N\right)}{N}} e^{j 2 \pi \frac{\left(l-l_p-m N\right)\left(k_p+n M\right)}{M N}} \nonumber \\ 
 & & w\left[k-k_p-n M, l-l_p-m N\right].
\end{eqnarray}
In order to make $|x_{_{\mbox{\scriptsize{s,dd}}}}[k,l]|$ almost constant for all $(k,l)$, $M$ and $N$ should be odd primes and $u$ should be relatively prime to both $M$ and $N$. Also $\sum_{k=0}^{M-1}\sum_{l=0}^{N-1} |x_{\mathrm{s}, \mathrm{dd}}[k, l] |^2 = 1$. Also, other spread pilots, like ZC pilots defined in \cite{Mattu_ZC} can be used. 
\section{Sub-THz Testbed Overview}
\label{section: testbed overview}
Transmitting within the \gls{sub-thz} band requires the use of specially designed \gls{tx} and \gls{rx} modules. We utilize two upconverter/downconverter pairs, one which operates at a center frequency of 140~GHz and another at 240~GHz. The 140~GHz frontends are custom-built \gls{mixamc} units from \gls{vdi}.  The \gls{mixamc} has a multiplication factor of 12 and a WR-6.5 conical horn antenna with 13$^\circ$ half-power beamwidth. The 240~GHz frontends are commercial off-the-shelf compact converter up/down converters (CCU and CCD). The CCU and CCD units have a 6$\times$ multiplication factor and WR-4.3 conical horn antennas with 13 $^\circ$ half-power beamwidth. Starting with the transmit side, the upconverter module is driven by an external \gls{lo} signal provided by Keysight \gls{psg}. The \gls{psg} is a low-noise signal generator that operates up to 40~GHz, which provides a high enough frequency to drive our frontend systems given their multiplication factors. The \gls{lo} gets multiplied and mixed with an \gls{if} signal. In our case, this \gls{if} signal is our \gls{otfs} signal provided by a \gls{usrp}. \gls{usrp} B-210 models generate signals between 70~MHz up to 6~GHz and has an internal clock rate of up to 60~MHz. On the receive side, we similarly, connect a downconverter frontend to a \gls{psg} to supply the \gls{lo} and a \gls{usrp} for signal processing.  Both transmit and receive \gls{usrp}s are controlled by a laptop running MATLAB commands for automated control, leveraging Communications Toolbox Support Package for \gls{usrp} Radio. Our testbed setup is shown in Fig.~\ref{fig:testbed} and the connection diagram in Fig.~\ref{fig:connection diagram}. 
\section{\gls{otfs} Implementation}
\label{section:Otfs implementation}
To complete over-the-air experimentation, we must synthesize our signal processing and design with the wireless testbed hardware and  testing workflow. In this section, we detail our implementation and define the parameters of the experiment. \vspace{-2mm}
\subsection{Baseband Signal Design \& Processing} \label{baseband_sd}
\par The \gls{td} transmit signals are packet-like, containing a header sequence for synchronization and recovery followed by the baseband Zak-\gls{otfs} signal.
We used a three-segmented Zadoff-Chu sequence with different lengths as the header sequence to synchronize data and estimate \gls{cfo} \cite{Zhihui}.
For the experiments in this paper, we designed two  different baseband Zak-\gls{otfs} signals, one for point pilot case and the other for spread pilot case. 
\par For the point pilot case, we transmit pilot and data in separate \gls{dd} frames. In one \gls{dd} frame, we transmit only point pilot  at location $(k_p=\frac{M}{2},l_p=\frac{N}{2})$ making other \gls{dd} resources as zero and in the adjacent frame, we transmit the information symbols. We separately transform these \gls{dd} frames into \gls{td} using Section \ref{subsection:zak_tx}. We concatenate these two \gls{td} signals to create a baseband Zak-\gls{otfs} signal ready to sense and communicate. At the receiver, once syncronized, we deconcatenate and transform separately the pilot and data signal from \gls{td} to \gls{dd} domain. The predicted \gls{dd} channel from the received \gls{dd} pilot signal is used to equalize the data in the data frame\footnote{For the setup, the channel remains constant for the two \gls{otfs} frames.}. 
\par For the spread pilot case, we use (\ref{spreadpilot_eqn}) and $(k_p = \frac{M+1}{2},l_p=\frac{N+1}{2})$\footnote{$M, N$ are even for point pilot case and $M, N$ odd for spread pilot case.} to create a \gls{dd} spread signal which is added to the \gls{dd} information symbols, and then the combined signal is transformed into \gls{td} to get the baseband Zak-\gls{otfs} signal \cite{ubadah2024}. Once syncronized at the receiver, we transform the \gls{td} signal to \gls{dd} domain and use the combined, spread pilot plus data, signal to predict the channel. 
Now using the predicted channel, we remove the contribution of the spread pilot from the received \gls{dd} signal to get the received \gls{dd} data frame. 
\par In both cases, point pilot and spread pilot, we predicted the \gls{dd} channel by the cross-ambiguity operation between the received \gls{dd} signal and the pilot signal. There are different energies associated with each \gls{dd} information signal and the pilot signals\footnote{More details on cross-ambiguity and the energies of pilot and data signals are given in \cite{ubadah2024}} \cite{ubadah2024}.
\color{black}


\vspace{-2mm}
\subsection{Over-the-air Communication}
The wireless transmission workflow is as shown in Fig. \ref{fig:connection diagram}. We generate the \gls{td} transmit signal, from Section \ref{subsection:zak_tx}, using our custom-built Zak-\gls{otfs} transmit MATLAB code. 
This signal is sent from the computer to the transmit side \gls{usrp} via USB and MATLAB commands. The \gls{usrp} converts the input signal from baseband to \gls{rf} and outputs an \gls{rf} signal at the specified center frequency. This signal serves as the input of the \gls{if} signal to the \gls{tx} frontend. We set the \gls{lo} to the appropriate frequency based on the multiplier and center frequency (approximately 11.7 GHz for the \gls{mixamc} and 40~GHz for the CCU/CCD units) of the \gls{tx} frontend system. Driven by the \gls{lo} and \gls{if} signals, the \gls{tx} frontend upconverts this signal to its associated sub-THz frequency (140~GHz for the \gls{mixamc}s and 240~GHz for the CCU/CCD systems). The \gls{tx} frontend transmits the upconverted signal over-the air to the \gls{rx} frontend. Taking the input \gls{lo}, the \gls{rx} frontend downconverts the received signal and outputs to the receive-side \gls{usrp} (See Fig. \ref{fig:connection diagram}) . The \gls{usrp} downconverts the \gls{if} signal to baseband and interfaces with the computer for receive processing. The signal processing described in Section \ref{subsection:receiver} is handled by the custom build Zak-\gls{otfs} receive MATLAB code.  \vspace{-2mm}
\subsection{Zak-\gls{otfs} Experiments}
To showcase proof-of-concept transmission and reception, we transmitted the two different baseband Zak-\gls{otfs} signals (discussed in Section \ref{baseband_sd}) over-the-air on the testbed with both the 140~GHz and 240~GHz frontends. We set the distance between the \gls{tx} and \gls{rx} to a fixed value of 1~m and varied the power by adjusting the internal gain on the \gls{usrp}; this varies the \gls{snr} scenarios under test. For the Zak-\gls{otfs} testing with point pilot, we varied the gain between 40-60~dB to collect results for different \gls{snr}s, to gain insight on the performance in different channel conditions (noisy versus clear). For the spread pilot \gls{otfs} case, we test for a set \gls{usrp} gain 60~dB and look for the channel prediction along with recovering the overlayed data. Given the stationary, \gls{los}, set up with no know interferers, we expect a strong \gls{los} component with no reflections. Also,
the design of the \gls{dd} grid is tailored less to the expected channel conditions and more to the hardware limitations. \vspace{-3mm}
\section{Results}
\label{section:results}
\begin{figure}[htb]
    \centering
    \includegraphics[width=0.9\linewidth]{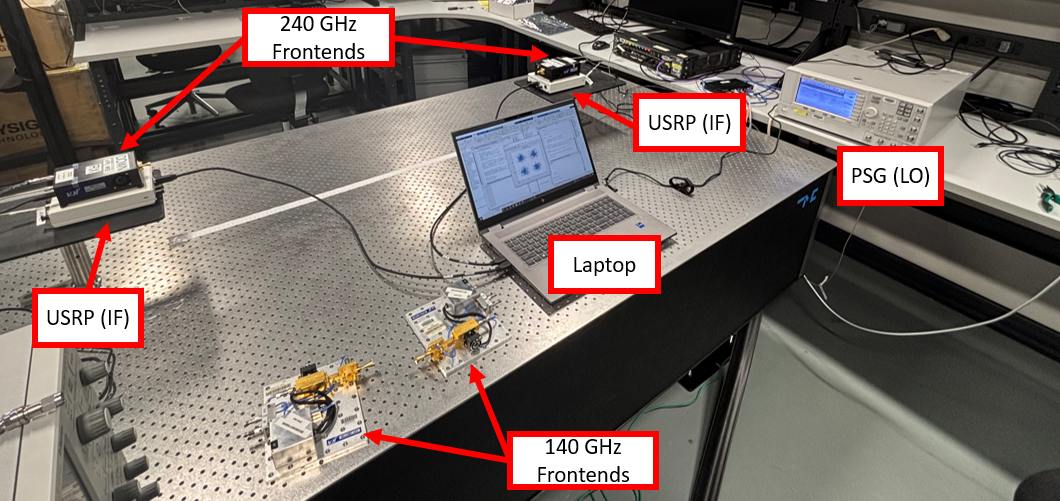}
    \caption{\gls{otfs} on \gls{sub-thz} laboratory set-up}
    \label{fig:testbed}
\end{figure}

 The THz testbed used for the over-the-air experimentation of Zak-\gls{otfs} waveform is shown in Fig.~\ref{fig:testbed}. The different parameters, both hardware and the Zak-\gls{otfs} system parameters, used or the experiments on the THz testbed, are shown in Table~\ref{parameters}. In this section, we present the results for both data transmission and sensing capabilities of Zak-\gls{otfs} system at THz frequencies. \vspace{-2mm}  
\subsection{Zak-\gls{otfs} results with point pilot}
\begin{figure}[htb]
    \centering
      \includegraphics[width=\linewidth]{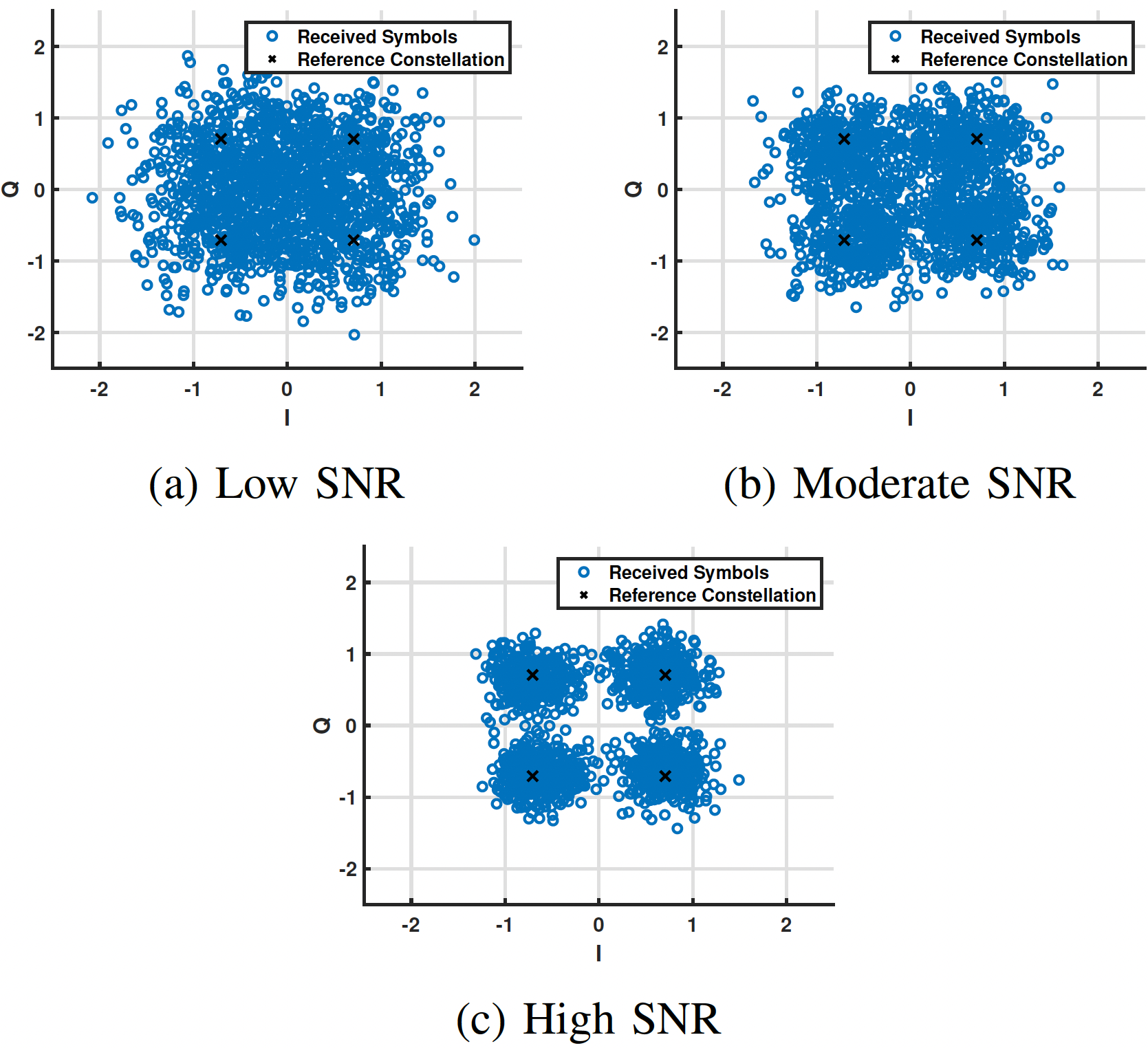}
    \caption{Zak-\gls{otfs} constellations at 140~GHz}
    \label{fig:zak results 140}
\end{figure}

\begin{figure} [htb]
    \centering
   \includegraphics[width=\linewidth]{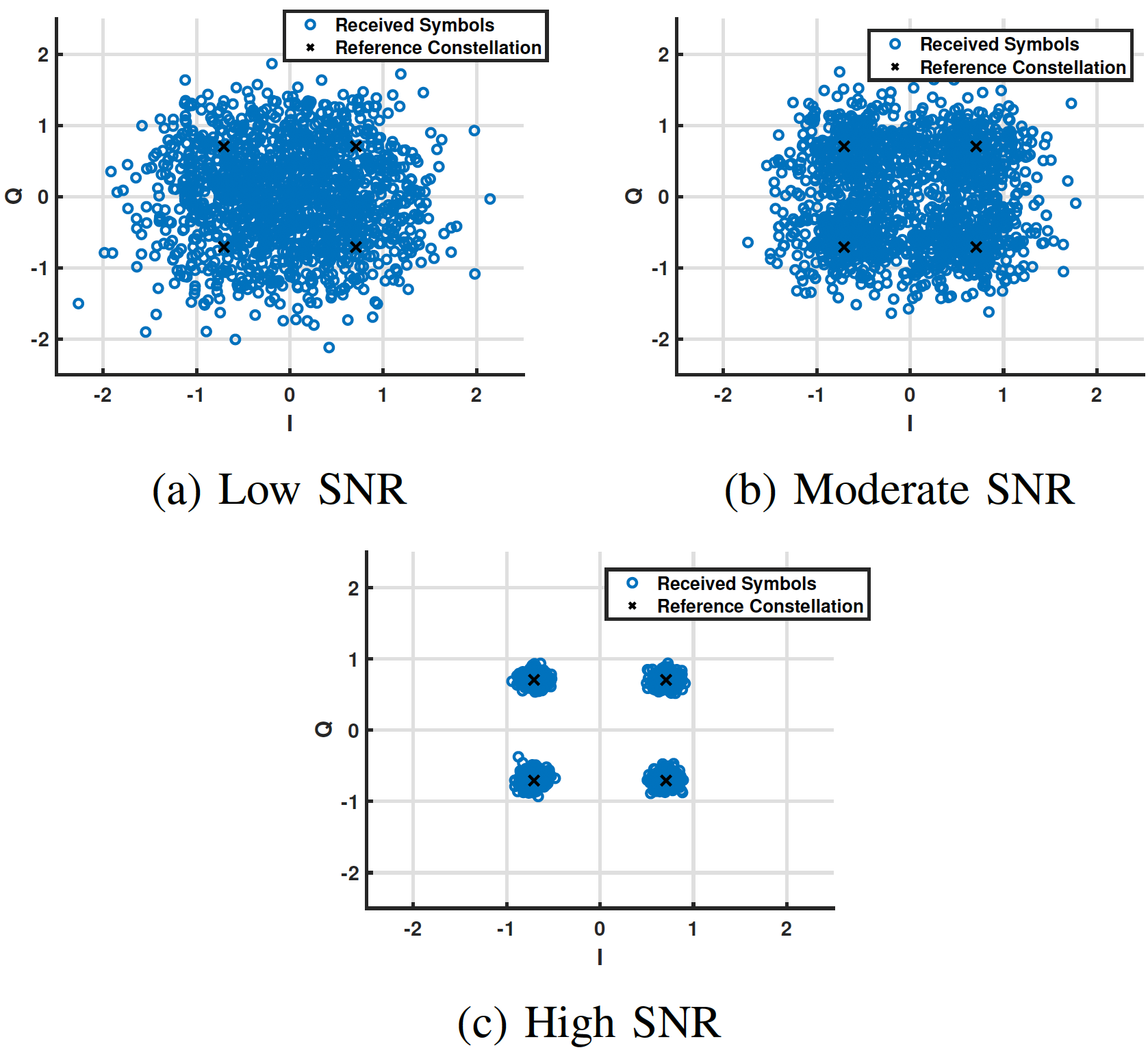}
    \caption{Zak-\gls{otfs} constellations at 240~GHz}
    \label{fig:zak results 240}
\end{figure}

We successfully transmitted the Zak-\gls{otfs} waveform with point pilot described in Section \ref{section:Otfs implementation} at 140~GHz and 240~GHz \gls{rf} frontends in three \gls{snr} regions: ``low" ($\leq$8~dB), ``moderate" ($\approx$12~dB), ``high" ($\geq$20~dB). Fig.~\ref{fig:zak results 140} shows the received constellations at 140~GHz for the three SNR regions described above. The \gls{ber} obtained in this three SNR regions are: 0.13 at 7.6~dB (low) , 0.05 at 11.6~dB (moderate), and $9.7\times10^{-4}$ at 21.9~dB (high). 
\begin{table}[htbp]
\caption{Zak-\gls{otfs} \& Hardware Parameters}
\begin{center}
\vspace{-3mm}
\begin{tabular}{|c|c|c|}
\hline
Parameters & Point Pilot Case & Spread Pilot Case \\
\hline
Base Modulation & QPSK & QPSK\\
Doppler Period ($\nu_p$) & 30~kHz & 30~kHz\\
Delay Period ($\tau_p = 1/\nu_p$) & 33.3 $\mu$s & 33.3 $\mu$s\\
Doppler Taps ($N$) & 48 & 37\\
Delay Taps ($M$) & 32 & 31\\
Bandwidth ($B$) & 960~kHz & 930~kHz\\
USRP Tx/Rx Gain & Varied & Varied\\
Tx/Rx Separation & 1~m & 1~m\\
IF Frequency & 2~GHz & 2~GHz\\
RF Frequencies & 140, 240~GHz & 140, 240~GHz\\
\hline
\end{tabular}
\label{parameters}
\end{center}
\vspace{-5mm}
\end{table} 
Also, in Fig.~\ref{fig:zak results 240}, we show the received constellations at 240~GHz for the same three SNR regions and the correspondinf \gls{ber}s are: 0.16 at 6.8~dB (low) , 0.05 at 12.4~dB (moderate), and no errors at 29.0~dB (high). Of note, our ability to access higher \gls{snr} regions with the 140~GHz system is limited by the additional multiplier stage in the hardware compared to the 240~GHz frontends; each multiplier is an additional source of noise, so 140~GHz captures, in-general, tend to be noisier than 240~GHz for the same input power. Therefore, our high-\gls{snr} case for 140~GHz is at a lower \gls{snr} than the 240~GHz system and has noisier results. 


\vspace{-6mm}

\vspace{-5mm}
\subsection{Zak-\gls{otfs} results with the spread pilot}
\begin{figure}[htb]
    \centering
    \begin{subfigure}{.45\linewidth}
    \includegraphics[width=\linewidth]{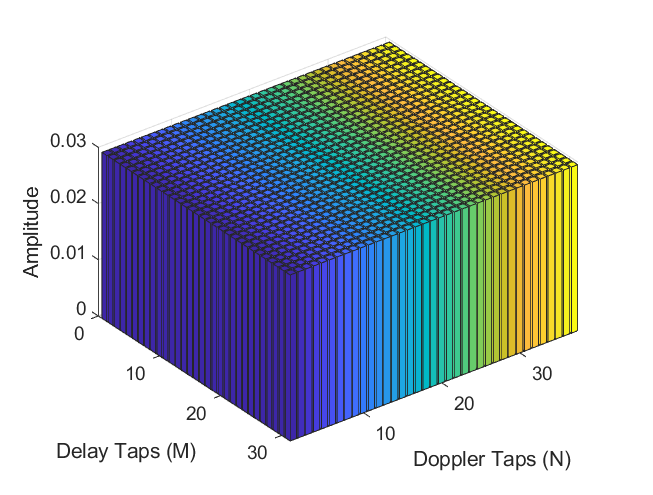}
        \caption{Transmitted pilot}
    \end{subfigure}
    \\
    \begin{subfigure}{.45\linewidth}
        \includegraphics[width=\linewidth]{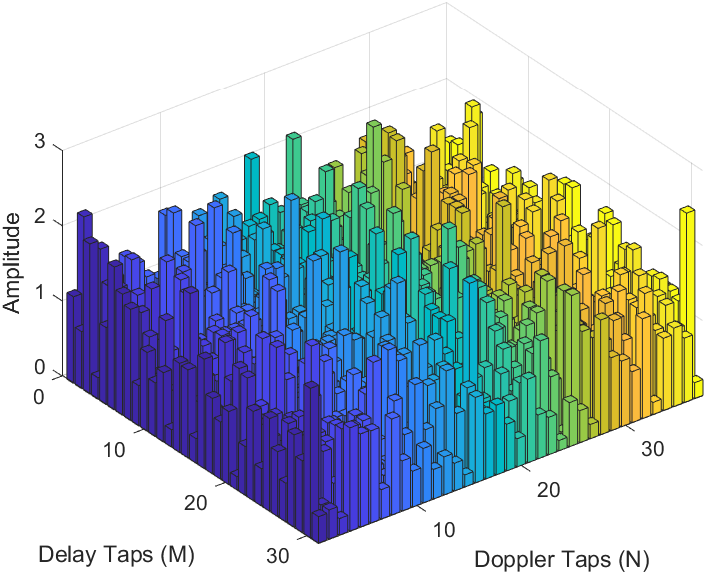}
        \caption{Received pilot 140 GHz}
    \end{subfigure}
    \begin{subfigure}{.45\linewidth}
        \includegraphics[width=\linewidth]{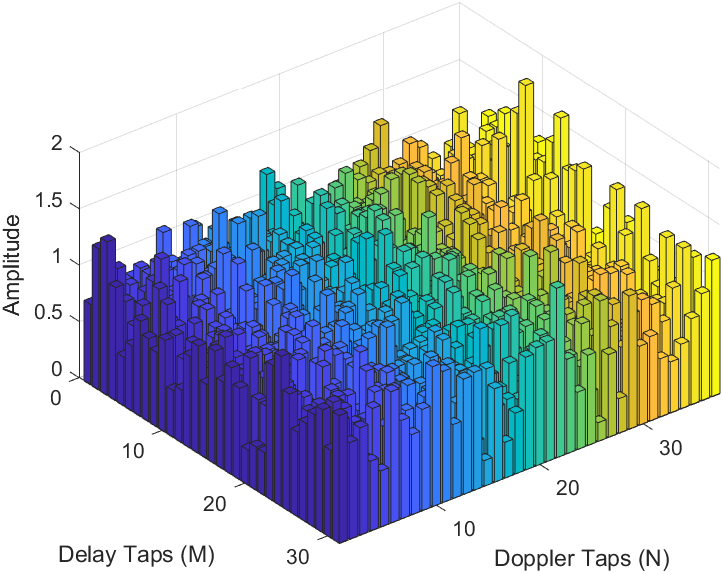}
        \caption{Received pilot 240 GHz}
    \end{subfigure}
    \\
    \begin{subfigure}{.45\linewidth}
        \includegraphics[width=\linewidth]{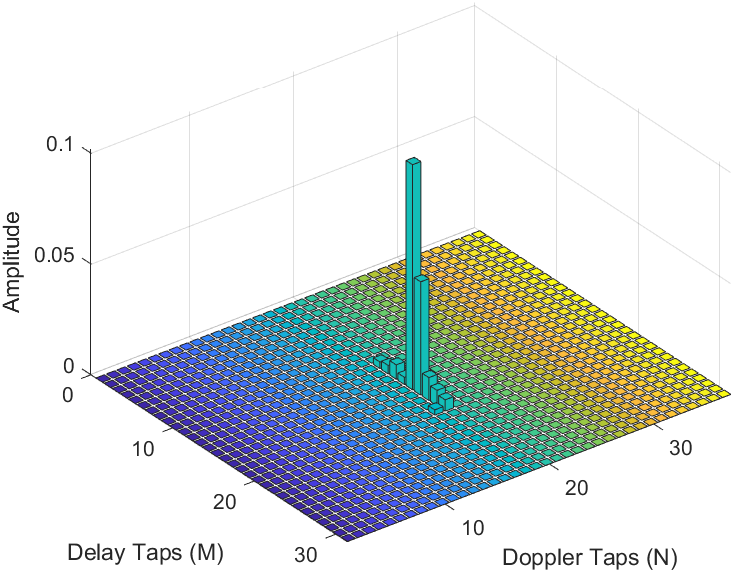}
        \caption{Predicted Channel at 140 GHz} 
    \end{subfigure}
        \begin{subfigure}{.45\linewidth}
        \includegraphics[width=\linewidth]{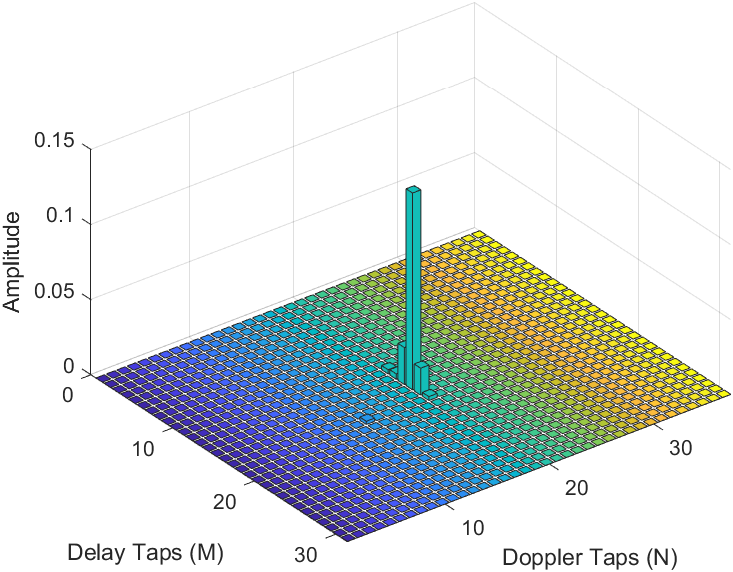}
        \caption{Predicted Channel at 240 GHz}
    \end{subfigure}
    \\
    \begin{subfigure}{.45\linewidth}
\includegraphics[width=\linewidth]{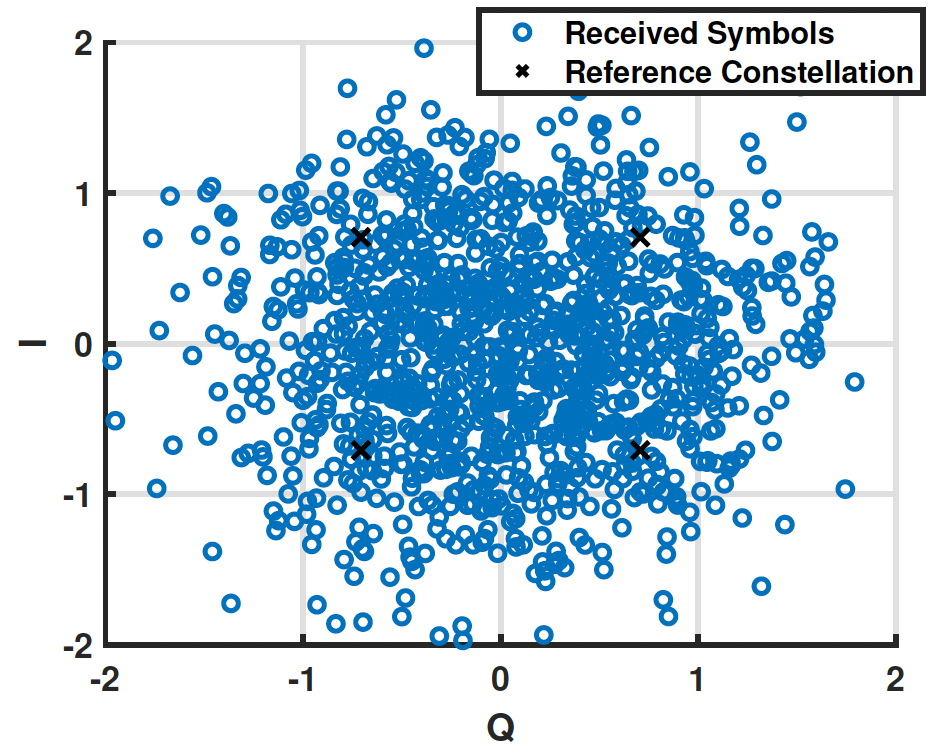}
        \caption{Recovered data 140 GHz}
    \end{subfigure}
    \begin{subfigure}{.45\linewidth}
    \includegraphics[width=\linewidth]{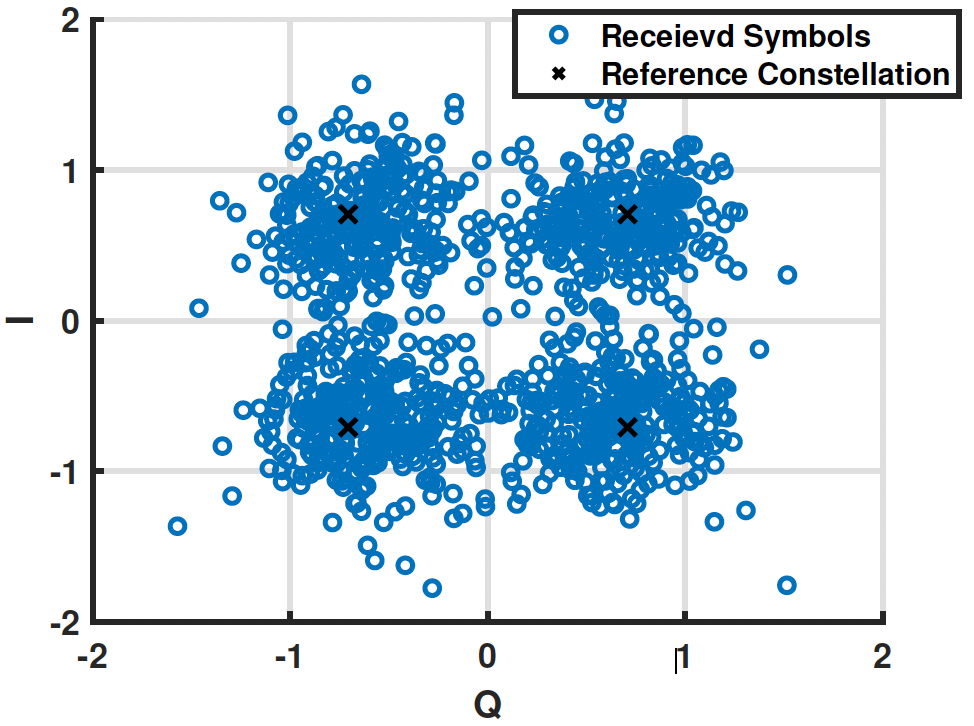}
        \caption{Recovered data 240 GHz}
    \end{subfigure}
    \caption{\gls{isac} for Zak-\gls{otfs} with spread pilots.}
    \label{fig:spreadpilot}
\end{figure}

As mentioned in Section~\ref{subsection:receiver}, to keep the absolute value of the spread pilot constant in all \gls{dd} resources, we choose $u=5$ coprime for both $M$ and $N$ (check Table \ref{parameters}). With this $u=5$ and the parameters from Table \ref{parameters}, we plot $|x_{_{\mbox{\scriptsize{s,dd}}}}[k,l]|$ in Fig.~\ref{fig:spreadpilot}(a) for the spread pilot. It can be seen in Fig.~\ref{fig:spreadpilot}(a) that all the \gls{dd} resources have equal/constant spread. Using the signal processing for spread pilot, mentioned in Section~\ref{section:Otfs implementation}, the recovered \gls{dd} signals at 140~GHz and 240~GHz are shown in Fig.~\ref{fig:spreadpilot}(b) and Fig.~\ref{fig:spreadpilot}(c) respectively. Note that the recovered \gls{dd} signals, Fig.~\ref{fig:spreadpilot}(b) and Fig.~\ref{fig:spreadpilot}(c), have data combined with the spread pilot. Using the cross-ambiguity operation, (refer to \cite{ubadah2024} for details), we first predict the channel response at 140~GHz and 240~GHz using Fig.~\ref{fig:spreadpilot}(b) and Fig.~\ref{fig:spreadpilot}(c) respectively. The predicted channel responses at 140~GHz and 240~GHz are shown in Fig.~\ref{fig:spreadpilot}(d) and Fig.~\ref{fig:spreadpilot}(e) respectively. Using this predicted channel, after removing the contribution from the spread pilot, we equalize the effect of the channel on the data. The recovered constellations at 140~GHz and 240~GHz are shown in Fig.~\ref{fig:spreadpilot}(f) and Fig.~\ref{fig:spreadpilot}(g) respectively. Focusing primarily on the channel prediction result, we observe a strong component centered within the \gls{dd} grid, with no other significant components. As expected for a stationary \gls{los} set-up, there is no \gls{dd} shift of the main component into another bin. No observed additional components showcases the lack of reflectors in our setup. Furthermore, we successfully demonstrated data recovery from the \gls{isac} frame. We note a \gls{ber} of 0.23 for 140~GHz case and \gls{ber} of 0.01 for 240~GHz (as showcased by the constellations in Fig.~\ref{fig:spreadpilot}(f) and Fig.~\ref{fig:spreadpilot}(g) respectively). We conducted this with the same \gls{usrp} gain as our high \gls{snr} scenario (60~dB). and find that there are some performance degradations from intertwining the pilot and data within the same frame, however, this can be improved with better decoding.
\vspace{-3mm}
In addition to the sensing capability, we measured the \gls{papr} of the point pilot compared to the spread pilot and found a performance improvement of $\approx$9~dB by using the spread pilot signal ($\approx$15~dB compared to $\approx$6~dB), which corroborates the results presented in~\cite{ubadah2024}.  For the entire signal (pilot and data together), we observe $\approx$11~dB \gls{papr} for the point-pilot Zak-\gls{otfs} signal compared to the $\approx$6~dB with the spread pilot, which makes $\approx$5~dB improvement. For \gls{sub-thz} systems, which are very sensitive to \gls{papr}, this is a promising result. 
\section{Conclusion}
\label{section:conclusion}
In this paper, we presented Zak-\gls{otfs} with point and spread pilots as potential candidates for next-generation \gls{sub-thz} communication systems. We demonstrated both waveforms over-the-air on practical \gls{sub-thz} wireless communications testbed and evaluated data recovery in a variety of \gls{snr} conditions. In addition, we showcased \gls{isac} capabilities and \gls{papr} improvements with spread-pilots. Future work includes more comprehensive \gls{ber} and \gls{snr} measurements to provide a true characterization of performance, as well as tailoring \gls{otfs} \gls{dd} parameters for ultra-wide bandwidths enabled by the \gls{sub-thz} spectrum.  

\section*{Acknowledgment}
The Duke team is supported by the US National Science Foundation (NSF) (grants 2342690,  2342690, and 214821), in-part by the Air Force Office of Scientific Research (grants FA8750-20-2-0504 and FA9550-23-1-0249), and in-part by federal agency and industry partner funds as specified in the Resilient \& Intelligent NextG Systems (RINGS) program. We thank Dr. Saif Khan Mohammed, Jinu Jayachandran from IIT Delhi, India, and Dr. Sandesh Rao Mattu from Duke University, USA, for helping us in this work.
Approved for Public Release; Distribution Unlimited: AFRL-2025-1371.
\bibliographystyle{ieeetr}\vspace{3mm}
\bibliography{References}

\begin{thebibliography}{10}

\bibitem{Elayan2020}
H.~Elayan, O.~Amin, B.~Shihada, R.~M. Shubair, and M.-S. Alouini, ``Terahertz band: The last piece of rf spectrum puzzle for communication systems,'' {\em IEEE Open Journal of the Communications Society}, vol.~1, pp.~1--32, 2020.

\bibitem{Jiang2024}
W.~Jiang, Q.~Zhou, J.~He, M.~A. Habibi, S.~Melnyk, M.~El-Absi, B.~Han, M.~D. Renzo, H.~D. Schotten, F.-L. Luo, T.~S. El-Bawab, M.~Juntti, M.~Debbah, and V.~C.~M. Leung, ``Terahertz communications and sensing for {6G} and beyond: A comprehensive review,'' {\em IEEE Communications Surveys \& Tutorials}, vol.~26, no.~4, pp.~2326--2381, 2024.

\bibitem{Akyildiz2022}
I.~F. Akyildiz, C.~Han, Z.~Hu, S.~Nie, and J.~M. Jornet, ``Terahertz band communication: An old problem revisited and research directions for the next decade,'' {\em IEEE Transactions on Communications}, vol.~70, no.~6, pp.~4250--4285, 2022.

\bibitem{Lee2020}
D.~Lee, A.~Davydov, B.~Mondal, G.~Xiong, G.~Morozov, and J.~Kim, ``From sub-terahertz to terahertz: challenges and design considerations,'' in {\em 2020 IEEE Wireless Communications and Networking Conference Workshops (WCNCW)}, pp.~1--8, 2020.

\bibitem{Hadani2017}
R.~Hadani, S.~Rakib, M.~Tsatsanis, A.~Monk, A.~J. Goldsmith, A.~F. Molisch, and R.~Calderbank, ``Orthogonal time frequency space modulation,'' in {\em 2017 IEEE Wireless Communications and Networking Conference (WCNC)}, pp.~1--6, 2017.

\bibitem{Taraboush2022}
S.~Tarboush, H.~Sarieddeen, M.-S. Alouini, and T.~Y. Al-Naffouri, ``Single- versus multicarrier terahertz-band communications: A comparative study,'' {\em IEEE Open Journal of the Communications Society}, vol.~3, pp.~1466--1486, 2022.

\bibitem{Parisi2024}
C.~T. Parisi, S.~Badran, P.~Sen, V.~Petrov, and J.~M. Jornet, ``Modulations for terahertz band communications: Joint analysis of phase noise impact and {PAPR} effects,'' {\em IEEE Open Journal of the Communications Society}, vol.~5, pp.~412--429, 2024.

\bibitem{Gaudio2022}
L.~Gaudio, G.~Colavolpe, and G.~Caire, ``{OTFS} vs. {OFDM} in the presence of sparsity: A fair comparison,'' {\em IEEE Transactions on Wireless Communications}, vol.~21, no.~6, pp.~4410--4423, 2022.

\bibitem{Liu2025}
Y.~Liu, M.~Chen, C.~Pan, T.~Gong, J.~Yuan, and J.~Wang, ``{OTFS} versus {OFDM}: Which is superior in multiuser {LEO} satellite communications,'' {\em IEEE Journal on Selected Areas in Communications}, vol.~43, no.~1, pp.~139--155, 2025.

\bibitem{Wiffen2018}
F.~Wiffen, L.~Sayer, M.~Z. Bocus, A.~Doufexi, and A.~Nix, ``Comparison of {OTFS} and {OFDM} in ray launched sub-6 ghz and mmwave line-of-sight mobility channels,'' in {\em 2018 IEEE 29th Annual International Symposium on Personal, Indoor and Mobile Radio Communications (PIMRC)}, pp.~73--79, 2018.

\bibitem{ubadah2024}
M.~Ubadah, S.~K. Mohammed, R.~Hadani, S.~Kons, A.~Chockalingam, and R.~Calderbank, ``Zak-{OTFS} for integration of sensing and communication,'' 2024.

\bibitem{Saif2_base}
S.~K. Mohammed, R.~Hadani, A.~Chockalingam, and R.~Calderbank, ``{OTFS}—a mathematical foundation for communication and radar sensing in the delay-doppler domain,'' {\em IEEE BITS the Information Theory Magazine}, vol.~2, no.~2, pp.~36--55, 2022.

\bibitem{Saif2_chan}
S.~K. Mohammed, R.~Hadani, A.~Chockalingam, and R.~Calderbank, ``{OTFS}—predictability in the delay-doppler domain and its value to communication and radar sensing,'' {\em IEEE BITS the Information Theory Magazine}, vol.~3, no.~2, pp.~7--31, 2023.

\bibitem{Jinu_Gauss}
J.~Jayachandran, R.~K. Jaiswal, S.~K. Mohammed, R.~Hadani, A.~Chockalingam, and R.~Calderbank, ``Zak-otfs: Pulse shaping and the tradeoff between time/bandwidth expansion and predictability,'' 2024.

\bibitem{Das_Gauss}
A.~Das, F.~Jesbin, and A.~Chockalingam, ``A gaussian-sinc pulse shaping filter for zak-otfs,'' 2025.

\bibitem{saif_book}
S.~K. Mohammed, R.~Hadani, and A.~Chockalingam, {\em OTFS Modulation: Theory and Applications}.
\newblock Wiley-IEEE Press, 2024.

\bibitem{Mattu_ZC}
S.~R. Mattu, I.~A. Khan, V.~Khammammetti, B.~Dabak, S.~K. Mohammed, K.~Narayanan, and R.~Calderbank, ``Delay-doppler signal processing with zadoff-chu sequences,'' 2024.

\bibitem{Zhihui}
Z.~Gao, Z.~Qi, and T.~Chen, ``Mambas: Maneuvering analog multi-user beamforming using an array of subarrays in mmwave networks,'' ACM MobiCom '24, (New York, NY, USA), p.~694–708, 2024.

\end{thebibliography}

\end{document}